\documentstyle[preprint,prl,aps]{revtex}
\begin{document}
\draft
\title{Superconducting Coherence in a Vortex Line Liquid;
   Simulations with Finite $\lambda$}
\author{Tao Chen and S. Teitel}
\address{Department of Physics and Astronomy, University of Rochester,
 Rochester, New York 14627}
\date{\today}
\maketitle
\begin{abstract}
We carry out simulations of a lattice London superconductor
in a magnetic field ${\bf B}$, with a finite magnetic
penetration length $\lambda$.  We find that superconducting
coherence parallel to ${\bf B}$ persists into the vortex line
liquid.  We argue that the length scale relevant to this
effect is $\Lambda=\phi_0^2/8\pi T_m$.
\end{abstract}
\narrowtext

In high $T_c$ superconductors, thermal fluctuations are
believed to melt the ground state vortex line lattice
at temperatures well below the mean field $H_{c2}$
line \cite{Nel,FFH}.
The resulting vortex line liquid has received intense
theoretical and experimental study.  In particular, recent ``flux transformer"
experiments \cite{Safar,Cruz} on YBCO
show that superconducting coherence parallel to the
applied magnetic field ${\bf B}$ exists over very long
length scales well into the vortex liquid.

To investigate the fluctuating vortex line system, numerical
simulations have been carried out.
These have either involved a
simplified model of vortex line interactions \cite{Ryu},
or have used
the approximation \cite{FXY,Car,Li} that the bare magnetic penetration length
$\lambda\to\infty$, so that the magnetic field ${\bf B}$
inside the superconductor is uniform.
Such $\lambda\to\infty$ simulations \cite{Li} show a sharp
transition within the vortex line liquid, corresponding to the onset
of coherence parallel to ${\bf B}$.  While this
$\lambda\to\infty$ model is suggested by the large values of
$\kappa=\lambda/\xi_0$ in the high $T_c$ materials, it may fail
close to $T_c$, where the correlation length may exceed
the finite bare $\lambda$; thus the finite $\lambda$ model
may display different critical behavior from the $\lambda\to\infty$
limit.  To investigate this possibility, we present here new
simulations of a system of fluctuating vortex lines, in which we include
the effect of magnetic screening on the vortex interactions \cite{Car2},
due to a finite $\lambda$.  We investigate the presence of
superconducting coherence within the vortex line liquid, and
discuss the length scale relevant for this effect.

Our model is a discretized lattice superconductor in the London
limit \cite{Gupta}.  For simplicity, we consider
isotropic couplings.  Following Carneiro {\it et al}. \cite{Car},
a duality transformation maps this model onto one of
interacting vortex lines, with Hamiltonian
%%%%%%%%%%%%% eH %%%%%%%%%%%%%%%%%%%%%%%%%%%%%%%%%%%%%%%%%%
\begin{equation}
   {\cal H}=2\pi^2J_0\lambda^2\sum_{i,j}{\bf n}({\bf r}_i)
   \cdot{\bf n}({\bf r}_j) G({\bf r}_i-{\bf r}_j).
\label{eH}
\end{equation}
%%%%%%%%%%%%%%%%%%%%%%%%%%%%%%%%%%%%%%%%%%%%%%%%%%%%%%%
Here $n_\alpha({\bf r}_i)$ ($\alpha=x$, $y$, $z$) is the integer vorticity
through plaquette $\alpha$ at site ${\bf r}_i$ of a cubic mesh of points,
and $G({\bf r})$ is the lattice London interaction, with Fourier
transform
%%%%%%%%%%%%% eG %%%%%%%%%%%%%%%%%%%%%%%%%%%%%%%%%%%%%%%%
\begin{equation}
   G_q=1/(1+\lambda^2Q^2),\quad
   Q^2\equiv \sum_{\mu=x,y,z}[2-2\cos q_\mu].
\label{eG}
\end{equation}
%%%%%%%%%%%%%%%%%%%%%%%%%%%%%%%%%%%%%%%%%%%%%%%%%%%%%%%
The coupling is $J_0=\phi_0^2\xi_0/(16\pi^3\lambda^2)$, with
$\phi_0$ the flux quantum, and $\xi_0$ the spacing of
the discrete mesh, which we identify with the bare coherence
length that determines the size of a vortex core.
Henceforth we measure lengths in units of $\xi_0$, and
temperature in units of $J_0$.
Periodic boundary conditions in all directions are taken.

To test for superconducting coherence, we consider
the helicity modulus $\Upsilon_\mu(q_\nu)$
(where $\mu\ne\nu$),
defined as the linear response coefficient giving the supercurrent
${\bf j}$ induced by a transverse perturbation in the vector potential
of the externally applied magnetic field,
$\delta A^{ext}_\mu(q_\nu){\bf\hat \mu}$,
%%%%%%%%%% ej %%%%%%%%%%%%
\begin{equation}
   j_\mu(q_\nu) =-\Upsilon_\mu(q_\nu)\delta A^{ext}_\mu(q_\nu).
\label{ej}
\end{equation}
%%%%%%%%%%%%%%%%%%%%%%

We have earlier derived \cite{Chen} an expression for $\Upsilon_\mu(q_\nu)$
in a continuum London superconductor.  The generalization to the
lattice superconductor is
%%%%%%%%%%%% eUps %%%%%%%%%%
%\begin{equation}
%\Upsilon_\mu({\bf q}))={[J_0\lambda^2] Q^2\over 1+\lambda^2Q^2}\left\{
%1-{4\pi^2J_0\lambda^2\over VT}{({\bf\hat\mu}\times{\bf\hat Q})_\alpha
%({\bf\hat\mu}\times{\bf\hat Q}^*)_\beta\langle n_\alpha({\bf q})
%n_\beta(-{\bf q})\rangle\over 1+\lambda^2 Q^2}\right\},
%\label{eUps}
%\end{equation}
%%%%%%%%%%%%%%%%%%%%%%
\begin{equation}
   \Upsilon_\mu(q_\nu)={[J_0\lambda^2] Q^2\over 1+\lambda^2Q^2}\left\{
   1-{4\pi^2J_0\lambda^2\over VT}{\langle n_\sigma(q_\nu) n_\sigma(-q_\nu)
   \rangle\over 1+\lambda^2 Q^2}\right\},
\label{eUps}
\end{equation}
%%%%%%%%%%%%%%%%%%%%%%
where $\mu$, $\nu$, $\sigma$
are a cyclic permutation of $x$, $y$, $z$, and $Q^2=2-2\cos q_\nu$
as $q_\mu$, $q_\sigma=0$.

In the absence of vortices, $\Upsilon_\mu( q_\nu)=[J_0\lambda^2]Q^2/
(1+\lambda^2Q^2)$ describes the total screening of the
Meissner state.  With vortices, an expansion
in powers of $Q^2$,
%%%%%%%%%%% enn %%%%%%%%%%
\begin{equation}
   \langle n_\sigma(q_\nu)n_\sigma(q_\nu)\rangle = n_0 +n_1Q^2+n_2Q^4+...,
\label{enn}
\end{equation}
%%%%%%%%%%%%%%%%%%%%&
leads to an expression for $\Upsilon_\mu(q_\nu)$ at small $q_\nu$,
in terms of
a renormalized coupling $[J\lambda^2]_{\mu R}$
and penetration length $\lambda_{\mu R}$,
%%%%%%%%%% eUpsR %%%%%%%%%%%
\begin{equation}
   \Upsilon_\mu(q_\nu)={[J\lambda^2]_{\mu R}\over 1+\lambda^2_{\mu R}Q^2}
\label{eUpsR}
\end{equation}
%%%%%%%%%%%%%%%%%%%%%
with
%%%%%%%% eJR %%%%%%%%%%%%
\begin{eqnarray}
   \gamma_\mu  & \equiv & {[J\lambda^2]_{\mu R}\over
   [J_0\lambda^2]} = 1-{4\pi^2J_0\lambda^2
   \over VT} n_0
\label{eJRa} \\
   {\lambda^2_{\mu R}\over\lambda^2} & = & 1+{4\pi^2J_0\over VT}
    {(n_1-n_0\lambda^2)\over \gamma_\mu}.
\label{eJRb}
\end{eqnarray}
%\begin{equation}
%   \gamma_\mu\equiv {[J\lambda^2]_{\mu R}\over
%   [J_0\lambda^2]}=1-{4\pi^2J_0\lambda^2
%   \over VT} n_0
%   \quad {\rm and}\quad {\lambda^2_{\mu R}\over\lambda^2}\equiv
%   1+{4\pi^2J_0\over VT} {(n_1-n_0\lambda^2)\over \gamma_\mu}.
%\label{eJR}
%\end{equation}
%%%%%%%%%%%%%%%%%%%%%

To see the meaning of $\gamma_\mu$ and $\lambda_{\mu R}$, note
that the current $j_\mu(q_\nu)$ produced by
$\delta A^{ext}_\mu(q_\nu)$, induces a magnetic
vector potential due to Ampere's Law,
which in our units is $[J_0\lambda^2]Q^2\delta A^{ind}_\mu(q_\nu)
=j_\mu(q_\nu)$.
The change in the total magnetic field due to the external perturbation
is therefore given by $\delta A^{tot}_\mu(q_\nu)=
\delta A^{ind}_\mu(q_\nu)+\delta A^{ext}_\mu(q_\nu)$, or
%%%%%%%% eAtot %%%%%%%%%%%%%%
\begin{equation}
   \delta A^{tot}_\mu(q_\nu)=\left[(1-\gamma_\mu)+\gamma_\mu {Q^2\over
   Q^2 +\lambda^{-2}_{\mu R}}\right]\delta A^{ext}_\mu(q_\nu).
\label{eAtot}
\end{equation}
%%%%%%%%%%%%%%%%%%%%%%%%
Thus a fraction $1-\gamma_\mu$
of $\delta A^{ext}_\mu(q_\nu)$ penetrates into the material;
the remaining fraction $\gamma_\mu$ is screened out,
on the length scale $\lambda_{\mu R}$.
Equivalently, $\lambda_{\mu R}$ is the
length on which fluctuations in magnetic field decay to
equilibrium.
For a perfect Meissner effect, $\gamma_\mu =1$, and $\lambda_{\mu R}$
agrees with the usual definition of the London penetration
length \cite{Baym}.
We therefore interpret $1/\lambda^2_{\mu R}\sim\rho_{s\mu}$
as the density of superconducting electron pairs in direction
${\bf\hat\mu}$, even in the more general case of a
partial Meissner effect in the mixed state.
Although $\Upsilon_\mu(q_\nu)$ will have the same form Eq.(\ref{eUpsR})
in both the superconducting and the normal metal state (with ordinary
fluctuation diamagnetism),
a transition will be signaled some singularity
in $\gamma_\mu$ and $\lambda_{\mu R}$.
We focus now on $\gamma_\mu$.

Consider a uniform applied ${\bf H}=H{\bf\hat z}$.
$\Upsilon_x(q_y)$ and $\Upsilon_y(q_z)$ then describe the response
to external fields
$\delta H_z(q_y)$ and $\delta H_x(q_z)$, which represent
compression and tilt perturbations of ${\bf H}$, respectively.
Correspondingly one finds \cite{Chen}
%%%%%%%% egxy %%%%%%%%%%
\begin{equation}
   1-\gamma_x=dB_z/dH_z \quad {\rm and}\quad
   1-\gamma_y=dB_x/dH_x,
\label{egxy}
\end{equation}
%%%%%%%%%%%%%%%%%%%%%
where these susceptibilities are evaluated
at the applied field ${\bf H}$.  Since the high $T_c$
materials display strong fluctuation diamagnetism even in the normal
state, it is unclear whether or not $\gamma_{x,y}$ will display a pronounced
feature at the superconducting transition.

For behavior along ${\bf\hat z}$ however, parallel to ${\bf H}$,
the criteria for a superconducting transition is more clearly
defined.  $\Upsilon_z(q_x)$, describes the response to an external
field $\delta H_y(q_x)$, representing a combined shear and tilt
perturbation of the uniform applied $H{\bf\hat z}$.  In
Ref. \cite{Chen} we showed that for the vortex line lattice
one has $\gamma_z=1$ and a perfect Meissner screening.  For
a normal vortex line liquid however, $\gamma_z=1-dB_x/dH_x\ll 1$.
Thus the transition to the normal state is signaled by a
discontinuous jump in $\gamma_z$ from unity.
Expressed in terms of the expansion
coefficients of Eq.(\ref{enn}), we have superconducting coherence
along the field provided $n_0=0$; for the
normal state, $n_0>0$.

For our Monte Carlo simulation, we start with a fixed density
$f=B/\phi_0$ of
straight vortex lines parallel to ${\bf \hat z}$,
giving the ground state configuration for an internal
field $B{\bf\hat z}$.
Following Carneiro {\it et al}. \cite{Car}, we update the system, heating
from the ground state, by
adding elementary closed vortex rings (a square ring of unit
area) with random orientation and position.
These excitations are accepted or rejected according to
the standard Metropolis algorithm.  This provides a complete
sampling of phase space for the vortex variables ${\bf n}({\bf r}_i)$,
subject to the constraints that vorticity is locally conserved,
and the average internal
field, ${\bf B}=(\phi_0/V)\sum_i{\bf n}({\bf r}_i)=f\phi_0{\bf \hat z}$, is
constant.

Our simulations are for the case $f=1/15$, whose ground state on a cubic mesh
is a close approximation to a perfect triangular lattice.
We choose $\lambda=5$, comparable to the
vortex line spacing, $a_v\equiv 1/\sqrt{f}=3.87$.
We study system sizes $L_\perp=30$ in the $xy$ plane, and  $L_z=15$ and $30$
parallel to ${\bf H}$.
Each data point is typically
the result of $5,000$ sweeps to equilibrate, followed by $8-16,000$
sweeps to compute averages, where each sweep refers to $L_\perp^2L_z$ attempts
to add an elementary vortex ring.

In Fig.\ref{f1} we show a sample of our data, plotting
$\langle n_y(q_x)n_y(-q_x)\rangle$ vs. $q_x$, for various $T$, and $L_z=30$.
Fitting to Eq.(\ref{enn}) through $O(Q^4)$ yields the solid curves,
and determines the parameters $n_0$ and $n_1$.
Eq.(\ref{eJRa}) then gives the couplings $\gamma_\mu$, which we
plot vs. $T$ in Fig.\ref{f2}.
We see that $\gamma_{x,y}$ decrease towards zero at $T_m\simeq 1.2$,
while $\gamma_z$ decreases at $T_c\simeq 2.0$.  We also show
our results for $L_z=15$.

In Fig.\ref{f3} we show intensity plots of vortex correlations within
the same plane perpendicular to ${\bf B}$,
%%%%%%%%% eSq %%%%%%%%%%%%%%
\begin{equation}
   S({\bf q}_\perp)=\sum_{z,r_\perp}e^{i{\bf q}_\perp\cdot{\bf r}_\perp}
   \langle n_z({\bf r}_\perp,z)n_z(0,z)\rangle,
\label{eSq}
\end{equation}
%%%%%%%%%%%%%%%%%%%%%%%%%%%%%
where ${\bf q_\perp} =(q_x,q_y)$,
for various $T$, and $L_z=30$.
Below $T_m$ we see sharp Bragg peaks of a vortex
line lattice.  Above $T_m$ we see behavior characteristic of a liquid.
$T_m$ is thus the melting transition.  Since the discrete mesh of our
simulation acts like a periodic pinning potential for vortices, $T_m$
also coincides with a depinning of the vortex lines.  We believe
that the drop in $\gamma_{x,y}$ at $T_m$ is more a result of this
depinning, rather than a direct result of melting.  We expect from
Eq.(\ref{egxy}) that
$\gamma_{x,y}=1-dB_x/dH_x$ is finite above $T_m$,  but this value
is too small for us to determine accurately.

With respect to coherence along ${\bf H}$, we expect a discontinuous
jump in $\gamma_z$ from unity to $1-dB_x/dH_x\approx 0$ at $T_c$.
The finite width of the decrease observed in our data
is a finite size effect; we see that the transition
sharpens as $L_z$ increases.
We therefore estimate $T_c\simeq 2.0$, well into the vortex line
liquid.  This is the main result of our simulations.

Recent flux transformer
experiments on YBCO \cite{Safar,Cruz} show that there is a
temperature ``$T_{th}$"
below which vortex line correlations parallel to ${\bf H}$ become
comparable to the thickness of the sample.  ``$T_{th}$" is clearly above
the ``$T_{irr}$" where resistivity transverse to ${\bf H}$ vanishes.
Resistivity parallel to ${\bf H}$ however appears \cite{Cruz}
to vanish at ``$T_{th}$".
A similar conclusion concerning vortex line correlations may be inferred
from the measurements of Ref. \onlinecite{Kwok}, where the onset of
pinning by twin grain boundaries is shown to occur distinctly above a
sharp first order melting transition.  If we identify ``$T_{th}$" with
our $T_c$, and ``$T_{irr}$" with our $T_m$, our results are in complete
accord with these experimental findings.

We have also tried to compute the lengths
$\lambda^2_{\mu R}$, using Eq.(\ref{eJRb}) and our fitted $n_0$ and $n_1$.
However the factor
$(n_1-n_0\lambda^2)/\gamma_\mu$ that appears in Eq.(\ref{eJRb})
is, in the region of the transition, the quotient of
two small numbers each with large relative error.
We were therefore unable to obtain meaningful
results for $\lambda_{\mu R}$.

Although we have simulated with $L_z\gg\lambda$, one can still
question whether our results represent the true thermodynamic limit.
In particular, in Refs. \onlinecite{Safar,Cruz}, it was found that
``$T_{th}$" decreased towards ``$T_{irr}$" as sample thickness $L_z$
increased.
This is also consistent with our results for $\gamma_z$ in Fig.\ref{f2},
where there is some suggestion that, in addition to a sharpening of
the transition as $L_z$ increases, $T_c$ also decreases.
It is therefore important to note that there is another
length in the problem \cite{FFH}
%%%%%%%%%%% eLambda %%%%%%%%%%%%%
\begin{equation}
   \Lambda(T)\equiv \phi_0^2/(8\pi T) = (2\pi^2 J_0/T)\kappa\lambda.
   \label{eLambda}
\end{equation}
%%%%%%%%%%%%%%%%%%%%%%%%%%%%%%%%%
For our simulation, we have $30=L_z\ll\Lambda(T_m)\simeq 410$.

It has been argued \cite{Moore}
that $\Lambda$ determines the length on
which phase correlations
$C({\bf r})\equiv\langle e^{i[\theta({\bf r})-\theta(0)]}\rangle$
decay in the vortex line lattice;
however these same calculations show that in the vortex liquid,
the decay length  of $C({\bf r})$ is comparable to the
spacing between vortex lines $a_v\ll L_z$.  Thus this analysis of $C({\bf r})$
does not indicate why $\Lambda$ should be an important length
above $T_m$, where we continue to see superconducting coherence.

Another possibility is suggested by Nelson's analogy \cite{Nel} between
vortex lines and the imaginary time world lines of 2D bosons.  Nelson
argued that there should be a Kosterlitz-Thouless (KT) superfluid
transition of the analog bosons.
For $L_z$ sufficiently large, this
KT transition would occur at a
$T_{c}<T_m$, and so would be pre-empted by the formation of
the vortex line lattice.  But for $L_z$ small enough, $T_{c}>T_m$,
and one has a new state intermediate between the vortex line lattice
and the normal vortex line liquid \cite{noteFish}.  We now restate our
earlier calculation \cite{Chen}
of this $T_{c}$ in order to show that the length which
distinguishes between these two possibilities is $\Lambda(T_m)$.

As shown by Pollock and Ceperley \cite{Ceperley}, the 2D boson
superfluid density can be expressed in terms of the ``winding number"
${\bf W}$ which is the net spatial distance traveled by the ensemble
of bosons as they travel down the time axis of their world lines.
One has $\rho_s^{boson}=mT_{boson}\langle W^2\rangle/2\hbar^2$.
According to the KT theory, the 2D superfluid transition occurs at
a universal value of $\rho_s^{boson}$, which translates into the
condition $\langle W_{c}^2\rangle =4/\pi$.
In terms of vortex lines, ${\bf W}$ just measures the net vorticity
transverse to ${\bf H}$ \cite{Chen}.  We thus have \cite{noteYoung}
%%%%%%%%%% eW %%%%%%%%%%%%%%%
\begin{equation}
   \langle W_y^2\rangle = \lim_{q_x\to 0} {1\over L_\perp^2}
   \langle n_y(q_x)n_y(-q_x)\rangle
   \label{eW}
\end{equation}
%%%%%%%%%%%%%%%%%%%%%%%%%%%%%
This is precisely the same correlation as enters $\Upsilon_z$,
and comparing with Eq.(\ref{eJRa})
we can write $\gamma_z=1-(\Lambda(T)/L_z)\langle W_y^2\rangle$.
Thus $\rho_s^{boson}\propto\langle W^2\rangle =0$
implies $\gamma_z=1$; the normal boson fluid corresponds to a
vortex line liquid with coherence along ${\bf H}$.
$\rho_s^{boson}>0$ implies $\gamma_z<1$; the boson superfluid
corresponds to the normal vortex line liquid \cite{Feig}.

We have shown \cite{Chen} that for a normal vortex line liquid,
$\langle n_y(q_x)n_y(-q_x)\rangle=f^2L_\perp^2L_zT/c_{44}(q_x)$, where
$c_{44}(0)=(B^2/4\pi)dH_x/dB_x$ is the tilt modulus, and $f=B/\phi_0$.
Thus we conclude that $\langle W^2\rangle =\langle W_x^2 +W_y^2\rangle
=2\langle W_y^2\rangle = (8\pi L_zT/\phi_0^2)dB_x/dH_x$.  This
yields
%%%%%%%%% eKT %%%%%%%%%
\begin{equation}
   T_{c}={\phi_0^2\over 2\pi^2 L_z}{dH_x\over dB_x}, \quad {\rm or}
   \quad {T_{c}\over T_m}={4\over\pi}{\Lambda(T_m)\over L_z}
   {dH_x\over dB_x}.
\label{eKT}
\end{equation}
%%%%%%%%%%%%%%%%%%%%%%%
For large $B\gg H_{c1}$,
$dH_x/dB_x\simeq 1$.
Thus for $L_z<(4/\pi)\Lambda(T_m)$, one has $T_c>T_m$ and hence a vortex line
liquid with superconducting coherence along ${\bf H}$,
intermediate between the vortex line lattice, and the normal vortex
line liquid.  Only for $L_z>(4/\pi)\Lambda(T_m)$ will this intermediate state
disappear \cite{Feig2}.
For YBCO, with $T_m\simeq 90^\circ K$, one has
$\Lambda(T_m)\simeq 1400\mu m$, much thicker than
the samples ($\sim 50\mu m$) in
Refs.\onlinecite{Safar,Cruz,Kwok}.

We note that for  $B\gg H_{c1}$, $\Lambda(T_m)$ is a factor
$B/H_{c1}$ larger than the ``entanglement'' length originally
proposed by Nelson \cite{Nel} as the criterion for the 2D boson superfluid
transition.  This is because the notion of superfluidity, as
measured by ${\bf W}$, does not precisely correspond to
the geometric notion of line entanglement.  If just as many lines
wander to the right as wander to the left, one has ${\bf W}=0$, although
the lines may still be quite twisted and geometrically entangled.

Finally, we note that for our model, Eqs.(\ref{eLambda},\ref{eKT})
would predict
$T_c=(8\pi J_0 \kappa\lambda/L_z)T_m\simeq 25$, much higher than the
observed $T_c\simeq 2.0$.  We believe that this
results from a breakdown of the boson analogy near our observed $T_c$,
due to the proliferation of thermally excited
closed vortex rings, and intersections between vortex lines,
such as cause the transition in a $B=0$ model.
This is clearly the case for the $\lambda\to\infty$
model \cite{Li}, where $\lambda\to\infty$
at fixed $J_0$, also means $\Lambda\to\infty$.  Eq.(\ref{eKT})
would then imply $T_c\to\infty$.  The finite $T_c$ found in
$\lambda\to\infty$ simulations \cite{Li} therefore indicates
such a breakdown of the boson analogy.
Our present results concerning vortex ring distributions
and line intersections are similar to what was found for $\lambda\to\infty$
\cite{Li}.

To conclude, we find from simulations with $\lambda\ll L_z\ll\Lambda(T_m)$
(as is the case for recent experiments) that superconducting coherence
parallel to ${\bf H}$ persists into the vortex line liquid state.
We argue that this effect should vanish once $\Lambda(T_m)\le L_z$.

We would like to thank A. P. Young for very helpful discussions.
This work has been supported by DOE grant DE-FG02-89ER14017.

\bibliographystyle{unsrt}

\begin{figure}
\caption{$\langle n_x(q_y)n_x(-q_y)\rangle/L_\perp^2$ vs.
   $q_y=2\pi m/L_\perp$, ($m$ integer) for
   various $T$ and $L_\perp=L_z=30$. Sample error bars are shown.
   Solid lines are a fit to Eq.(\ref{enn}).}
\label{f1}
\end{figure}

\begin{figure}
\caption{$\gamma_\mu$ vs. $T$ for $L_\perp=30$ and $L_z=15,30$.
Sample error bars are shown.  $\gamma_{x,y}$ decreases
at $T_m\simeq 1.2$; $\gamma_z$ decreases at $T_c\simeq 2.0$.}
\label{f2}
\end{figure}

\begin{figure}
\caption{Intensity plot of $S({\bf q_\perp})$ for several $T$ and
$L_\perp=L_z=30$.
(a) $T=0.50<T_m$ shows a lattice of Bragg peaks; (b) $T=1.25 \approx
T_{m}$; (c) $T_m<T=1.60<T_c$ in the vortex line liquid; (d) $T=2.0>T_c$.}
\label{f3}
\end{figure}

\end{document}